\documentclass[12pt,preprint]{aastex}
\usepackage{emulateapj5,mathptmx}
\submitted{Submitted for Publication in The Astrophysical Journal}
\makeatletter

\newenvironment{inlinefigure}{%
\def\@captype{figure}%
\noindent\begin{minipage}{0.999\linewidth}\begin{center}}
{\end{center}\end{minipage}\smallskip}
\makeatother
\begin{document}

\title{A Compact Central Object in the Supernova Remnant Kes 79}

\author{F.D. Seward, P.O. Slane, R.K. Smith}

\affil{Smithsonian Astrophysical Observatory, Cambridge MA 02138}

\author{M. Sun}

\affil{Harvard University, Cambridge MA 02138}

\begin{abstract}
A Chandra X-ray observation has detected an unresolved source at the
center of the supernova remnant Kes 79.  The best single-model fit
to the source spectrum is a blackbody with an X-ray luminosity
$L_{X}$(0.3-8.0 keV)=$7\times 10^{33}$ ergs s$^{-1}$.  
There is no evidence for a
surrounding pulsar wind nebula.  There are no cataloged counterparts at
other wavelengths, but the absorption is high.  The source properties 
are similar to the central source in Cas A even though the Kes 79
remnant is considerably older.
\end{abstract}

\keywords{supernova remnants - stars:neutron - X-rays}

\section{Introduction}

Source 79 in the radio catalog of Kesteven (1968) lies directly in
the Galactic Plane, $33^{\circ}$ NE of the Galactic Center.  It is a
moderately large supernova remnant (SNR) and is sometimes called 
G33.6+0.1.  Radio observations (Velusamy, Becker \& Seward
1991) show an outer shell, $11^{\prime}$ in diameter, which is
approximately circular over the SW half of the remnant but with 
large indentations in the N and E boundaries.  The
brightest part of the 
radio remnant is an interior region with shell-like form.  The
southern part of this ``inner shell'' has the highest surface brightness
in the remnant.  The distance has been determined by neutral H
absorption measurements to be $10\pm2$ kpc by Frail \& Clifton (1989). 

The first X-ray detection of Kes 79 was made using the Einstein Observatory 
(Seaquist \& Gilmore 1982; Velusamy, Becker \& Seward 1991) and
showed amorphous  structure with a bright center.  This inspired a
ROSAT observation (Seward \& Velusamy 1995) to search for Crab-like
structure in the interior.  The results, however, showed no indication
of a central pulsar or pulsar-wind nebula.  The southern arc of the
inner ring was found to be bright in X-rays and faint 
emission was observed from the outer shell, particularly close to
the eastern indentation.  Assuming a thermal spectrum, analysis of the
ROSAT data indicated an age of $6-12\times 10^{3}$ years, an X-ray
luminosity of $\sim 10^{36}$ erg s$^{-1}$, and an energy
release of $5\times 10^{50}$ ergs in the SN explosion.  Subsequent
observations with ASCA (Sun \& Wang 2000) showed that the spectrum was
indeed thermal with strong lines from Mg, Si and S.  The global
spectrum was fit well by a single NEI model.  
Seward \& Velusamy (1995) speculated that this remnant might be
younger and closer than believed and the result of a Type Ia
supernova.  As evidence they cited: the circular shell, the fact that
the absorption was the same as that in the path to W44
($1^{\circ}$ distant in the plane of the sky and only 3 kpc distant
from the sun), and the lack of an observable pulsar.  The
Chandra observation described here shows that this is not the case.  

In this paper we report the detection of a point-like source at the
center of the remnant, which is likely to be a neutron star created in
the SN explosion. We discuss the spectrum, the apparent absence of any
surrounding synchrotron emission, and briefly compare the source
with similar objects.  Discussion of the SNR shell is defered to
a subsequent publication.

\section{Chandra Observation}

Chandra observed Kes 79 on 31 July 2001.  The observation was
undertaken to better determine the shell-like structure and to measure
spectra from different regions.  An exposure of 30 ks was obtained
and there were no ``flares'' from 
particle-induced background.  The remnant was centered in the ACIS-I
array and consequently spread over the four, $8^{\prime}$ square chips
comprising the detector.  The telescope was dithered and images are
exposure corrected so the gaps between chips do not appear in the
images.  Because of the CTI degradation of the
detector (Chandra X-ray Center 2001), the spectral resolution is
better along the NE and SW edges of the remnant than at the center.  
We took this into account using appropriate tasks from
the CIAO software, version 2.1, which was used for data analysis. 

\begin{inlinefigure}
\includegraphics[width=0.95\linewidth]{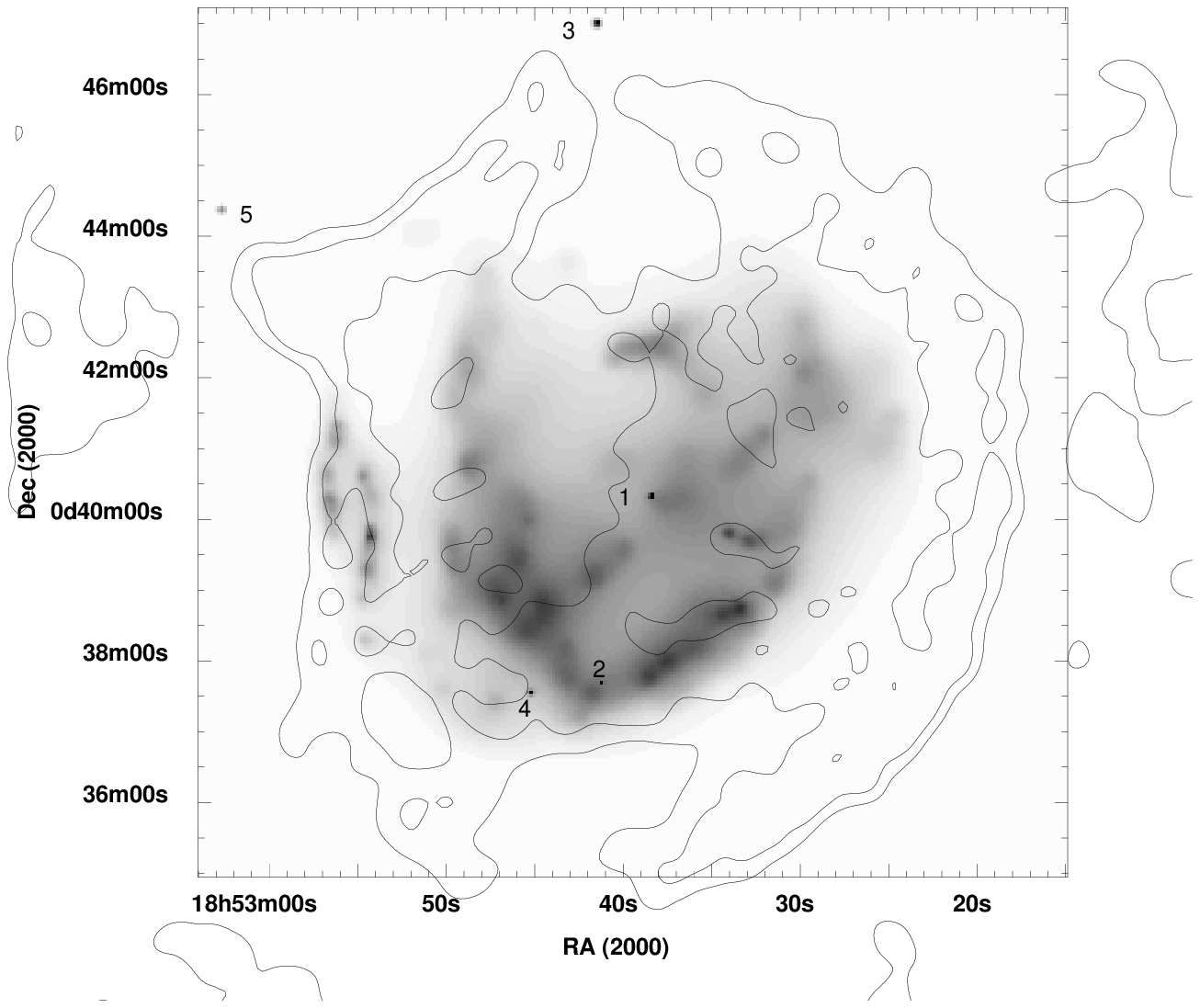}
\caption{Chandra image of Kes 79 in the energy range 0.8-8 keV.  Data have
been adaptively smoothed. 
The grey-scale X-ray emission comes largely from the bright inner ring
of the remnant, a region with dimensions 5$^{\prime}$ $\times
$6$^{\prime}$.  Unresolved sources are labeled 1-5.  Source 1
is centered in the remnant and is, by a factor of 10,
 the brightest point-like source in
this field.  The radio remnant is indicated by three brightness
contours at relative surface brightness levels of 1, 2.24, and 5.} 
\label{fig1}
\end{inlinefigure}

Figure 1 shows the Chandra image.  The data have been adaptively smoothed
(with the CSMOOTH algorithm) and show the bright inner shell of the
remnant.  Emission from the outer shell is weak and is hard to see
(except in the E indentation) in this figure.  Five unresolved sources
are easily seen, including 3 within the shell.  The brightest, by an
order of magnitude, is the source at the middle which is the subject
of this paper.  The sources are labeled 1-5 in Figure 1, and Table 1
lists their properties.  The May 2002 Chandra aspect solution was
used.  For sources within $2^{\prime}$ of the telescope axis the 90\% 
certainty radius in the Chandra position is $0.6^{\prime \prime}$. 
Positions were compared with the GSC 2.2 Catalog (STScI 2001).
Sources 3 ($7^{\prime}$ off-axis) and 4 ($3^{\prime}$ off-axis) 
are probably stars; the X-ray spectra are soft and the X-ray
source positions are $0.9^{\prime \prime}$ and $0.2^{\prime \prime}$
from optical counterparts, which shows the accuracy of the aspect
determination for these parts of the field.

Hardness ratios are given for eventual comparison with 2 surveys of
serendipitous Chandra sources, extragalactic (Green et al 2002) and
Galactic (Grindlay et al. 2002).   The hardness ratio, HR = H-S/H+S
where S is the number of counts from 0.3 to 2.5 keV and H is the
number of counts from 2.5 to 8 keV.  There is no counterpart brighter
than R magnitude 19 for the central source (1) or for sources 2 and 5.
Because the spectra of sources 2 and 5 are hard, they are probably 
background AGN.

The radial profile of the central source is consistent with the
expected Chandra point-spread-function.  Any extent is $<1.0^{\prime
\prime}$.  Thinking this likely to be an isolated neutron star, we
searched for time variability and pulsations.  The light curve is
constant within statistics - $2\sigma$ fluctuations $\leq$ 20\% on an
hourly basis.  The search for periodicity was limited to
periods longer than 6.4 s because the ACIS instrument, in normal mode,
integrates for 3.2 s.  An FFT analysis showed no coherent power significantly
above the noise, giving an upper limit of $\approx30$\% for pulsed
power in this range. 

\begin{inlinefigure}
\includegraphics[width=0.95\linewidth]{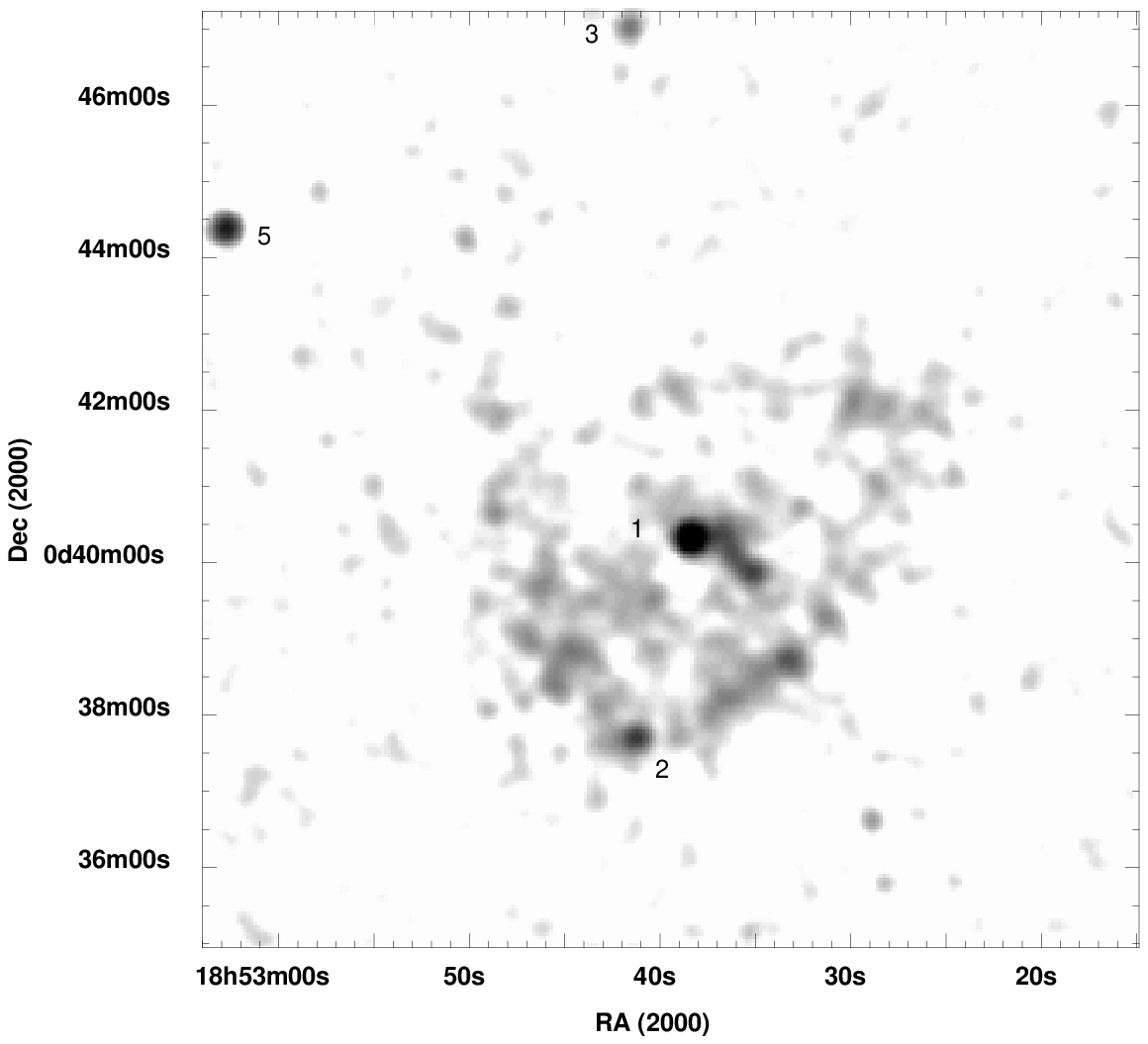}
\caption{Chandra image of Kes 79 in the energy range 3-5 keV. Data have
been smoothed with a Gaussian of FWHM=12$^{\prime \prime}$. Four of
the unresolved sources are still visible.}
\label{fig2}
\end{inlinefigure}

Since many young pulsars are embedded in diffuse synchrotron radiation
from a surrounding PWN, we searched for such a PWN at high
energies.  In the range 5-8 keV, the 
central source is still visible but diffuse emission is absent,
both from the bright inner shell and from the vicinity of the central
source.  In the range 3-5 keV, shown in Figure 2, the inner ring is
discernable and there is a small feature adjacent to the central
source and extending SW.  The spectrum of emission from this SW
feature shows Mg and Si lines, indicating that much of the emission is
thermal.  Since there are only 300
counts from this feature, model parameters cannot be accurately
determined from spectral fits.  By subtracting the thermal spectrum
observed elsewhere in the remnant we estimate that up to about 1/3 of the
emission from this  $\sim 0.6^{\prime} \times 0.6^{\prime}$
feature could be  non-thermal.  Assuming a power-law spectrum
with photon index = 2 (like the Crab Nebula), the upper
limit to the luminosity of a PWN at this location is calculated to 
be $1.5 \times 10^{33}$ ergs s$^{-1}$.  This upper limit is about
1/4 the luminosity of the point-like source.  We note that a PWN
is usually more luminous than the pulsar itself [20:1 for the
Crab Pulsar (Toor \& Seward 1977), 1.3:1 for the Vela Pulsar (Helfand et
al 2001)] 

The spectrum of the central source is shown in Figure 3 and Table 2
lists the results of spectral fits.  723 counts were extracted from
a circle of radius $4^{\prime\prime}$, grouped over 9 ACIS
energy channels, and were restricted to the energy band 0.8-4.7 keV.
The SHERPA software was used for spectral fitting.
A power-law gives a fit which, although it produces an acceptable
$\chi^2$ = 1.2, is obviously too weak in the range 2-4 keV and too
strong in the range 4-6 keV.  Furthermore, the
photon index, $\gamma = 4.2$, is higher than observed for most cosmic
sources and the value for absorption is high.  The ASCA measured
$N_{H}$ was $(1.75\pm.07)\times 10^{22}$ atoms  cm$^{-2}$ and
spectral fits to the bright thermal emission in our Chandra data yield $N_{H}
\approx 1.8\times 10^{22}$ atoms cm$^{-2}$.  The central source
spectrum is soft, softer than expected from an AGN ($\alpha
\approx 1.7$) or from a PWN ($\alpha \approx 1.5 - 2.5$)(e.g. Slane et
al 2000). 

\begin{inlinefigure}
  \includegraphics[width=0.95\linewidth]{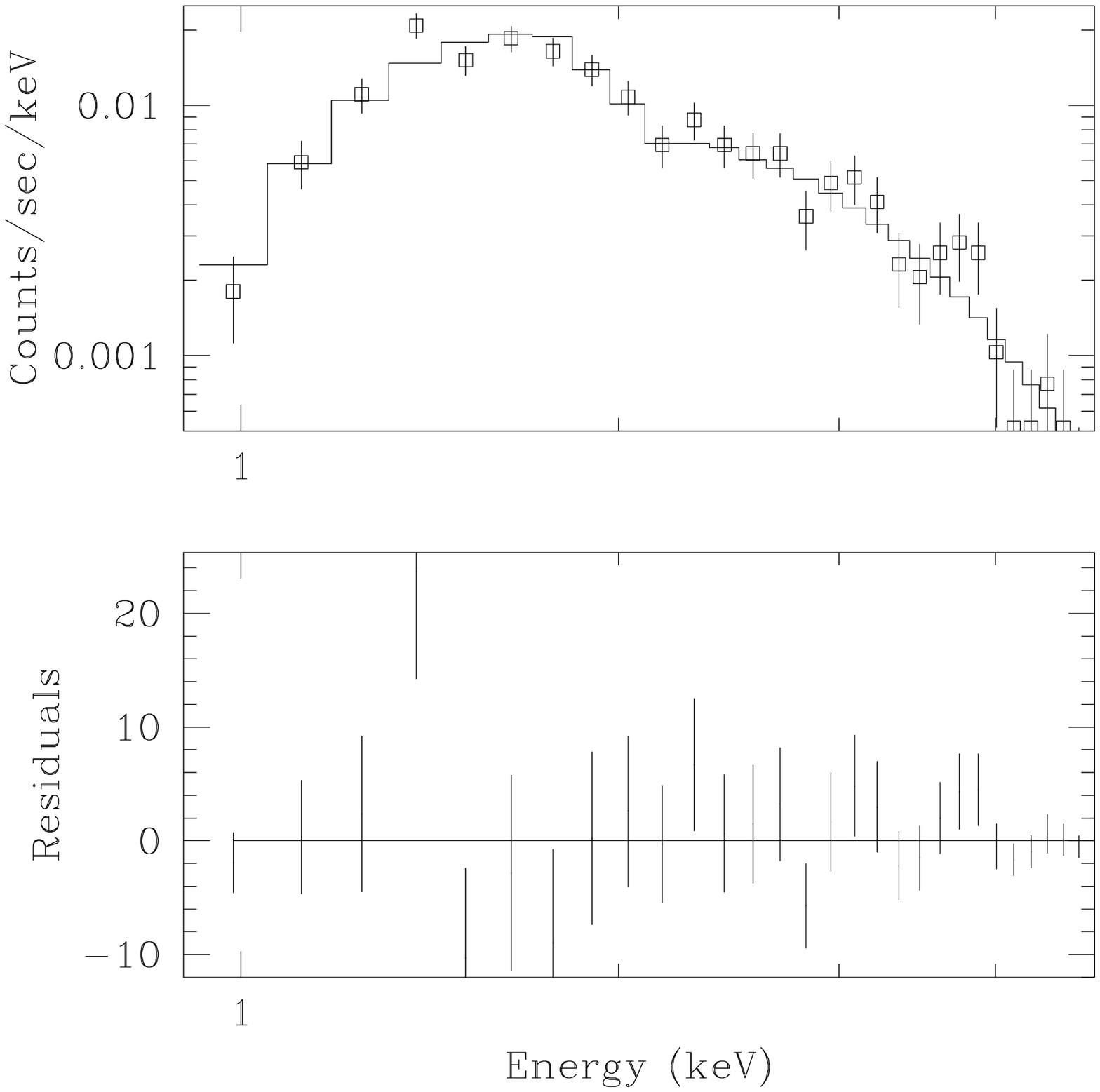}
\caption{X-ray spectrum of the central source.  A small background has
been subtracted.  The solid histogram is the result expected from a
black body spectrum with temperature kT = 0.48 keV. \label{fig3}}
\end{inlinefigure}

A single blackbody spectrum gives a good fit, shown in Figure 3,
but with $N_{H}$ a bit lower than expected. It is also not difficult to
achieve good fits with 2-component spectra.  
The soft component can be either blackbody or power law and is not
well constrained.  A power law does not work well as the hard 
component and, if used as the soft component, only makes a small
contribution to the emission.  
If we require that
$N_{H}=1.8\times 10^{22}$ atoms cm$^{-2}$, the absorption observed by
Chandra for the diffuse part of the remnant, then a 2-component
spectrum produces a better fit than a single blackbody spectrum but
the relative contribution of the soft component is small.
Values of parameters are given in Table 2.

Chandra spectra of various parts of the diffuse remnant are all
thermal.  The Mg, Si, and S lines detected by ASCA are prominent and
there is not much variation 
from region to region.  These data and their interpretation will be the
subject of a second paper (Sun et al. 2002). 

\begin{table*}
\tabletypesize{\small}
\begin{center}
\caption{Properties of Serendipitous Sources}
\begin{tabular}{lllcccl} \\ \hline\hline
 & & & ACIS & Hardness & & Optical\\
Source & RA & Dec & Counts & Ratio & R Mag & Counterpart \\ \hline
1, Central Source & 18h 52m 38.56s & +00$^{\circ}$ $40^{\prime}$
$19.84^{\prime \prime}$ &723 &$-0.49\pm0.04$ &$>18.5$&None \\
2 & 18h 52m 41.35s & +00$^{\circ}$ $37^{\prime}$ $42.5^{\prime
\prime}$ & \phantom{0}60 &
$\phantom{-}0.37\pm0.14$&$>18.5$&None \\
3, Star? & 18h 52m 41.66s & +00$^{\circ}$ $47^{\prime}$ $01.4^{\prime
\prime}$ & \phantom{0}61& $-0.39\pm0.19$&$\phantom{>}16.0$&N020120163554 \\
4, Star? & 18h 52m 45.38s & $+00^{\circ}$ $37^{\prime}$ $34.1^{\prime
\prime}$ & \phantom{0}47 & $-0.83\pm0.19$& $\phantom{>}16.3$&N020120158710 \\
5&18h 53m 02.87s&$+00^{\circ}$ $44^{\prime}$ $22.9^{\prime
\prime}$&\phantom{0}68& $\phantom{-}0.70\pm0.14$&$>18.5$& None \\ \hline
\end{tabular}
\end{center}
\end{table*}

The radio pulsar, B1849-00, proposed to be a high-velocity object
associated with Kes 79 (Han 1997), is located just outside
the remnant $3^{\prime}$ south of the rim.  Frail \& Clifton
(1989), however, obtained 21 cm absorption 
data which showed PSR 1849-00 to be considerably further away than Kes 79.
The present discovery of a different object at the center of Kes
79, verifies their conclusion that PSR 1849-00 is not associated 
with Kes 79. 

As a matter of interest, PSR 1849-00 was in the field of view of the
ACIS-I detector during our observation and was not detected.  An
upper limit on the flux is $1 \times 10^{-14}$ ergs cm$^{-2}$
s$^{-1}$.  If the pulsar were 15 kpc distant, and the transmission of the
ISM is $t_{ism}$, then $L_x$ would be $2.5 \times 10^{32}$
$t_{ism}$$^{-1}$ erg s$^{-1}$.  Since \.{E} is $\sim 4 \times 10^{32}$
erg s$^{-1}$ for this pulsar, $L_x$ is expected to be well below
this upper limit.

\section{Discussion}

The blackbody spectrum strongly suggests that the central point source
in Kes 79 is not a
foreground star or background AGN, as are the other unresolved sources
in Figure 1.  The central location and similarity to other
recently-discovered objects (see list 2 below) indicate that this is
probably a neutron star --- the remnant of the core of the star which
exploded to produce Kes 79.  It was not detected in previous
X-ray observations because the counting rate 
of the central source is only $\sim 10^{-2}$ that of the entire
remnant
in the 1-10 keV energy band.
The luminosity at 10 kpc distance, in the band 0.3-8 keV, is
$7\times 10^{33}$ erg s$^{-1}$, about 4 times the luminosity as that of the
central object in Cas A (Chakrabarty et al. 2001).
 The blackbody spectrum and lack
of a PWN favor thermal emission originating from a small region,
perhaps on the surface of the star.  

\begin{table*}
\begin{center}
\caption{Spectral Fits to the Central Source}
\begin{tabular}{llll}\\ \hline\hline
 & & & Reduced \\
Model & Spectral Index & $N_{H}$(10$^{22}$ atoms cm$^{-2}$) & $\chi^{2}$ \\
\hline
Power Law &$\alpha = 4.2\pm0.25$ &$2.7\pm0.2$ &1.19 (25 dof)\\
Black Body & $kT=0.48\pm0.025$ &$1.45\pm0.17$ &0.92 (25 dof)\\
2 Black Body & $kT=0.18\pm0.06$ &1.8 (fixed) &0.94 (24 dof)\\
& $kT=0.47\pm0.02$ \\ \hline
\end{tabular}
\end{center}
\end{table*}

Central objects in SNRs were well-understood when the only two known
were the Crab and Vela pulsars.  For each of these objects, the pulsar
characteristic age, P/2\.{P}, and the age of the surrounding remnant 
are about the same (950 years for the Crab and $\sim 10^4$ years for
the Vela Remnant), most of the energy radiated from the vicinity of
the pulsar is nonthermal, and spin-down energy is adequate to power all
emission from the pulsar and PWN.  Now that more objects have been
discovered, the situation is complex.

Some putative neutron stars within remnants show no sign of 
non-thermal emission.  There is
no PWN, the X-ray spectrum is more blackbody than power law, and no
radio or gamma-ray pulsations have been observed.  These have been called
``radio-quiet'' pulsars and/or CCOs (Compact Central Objects).
Such objects are found in Cas A (Murray et al. 2001; Chakrabarty et al. 2001),
 Pup A (Zavlin et al. 1999), PKS 1209-51/52 (Zavlin et al. 2000;
Pavlov et al. 2002) and G347.3-0.5 (Slane et al. 1999).
Properties of these objects are reviewed by Pavlov, Sanwal,
Garmire, and Zavlin (2002).  The Kes 79
source appears to be of this type.  All have luminosities, $L_{X}$
between $10^{33}$ and $10^{34}$ erg s$^{-1}$.
There is only one convincing case for pulsed
emission; the central source in PKS 1209-51/52 has a period of 0.424 s,
a sinusoidal pulse shape, and a pulsed fraction of
$\approx10$\% (Zavlin et al. 2000; Pavlov et al. 2002). 

We note that
Kes 79 was searched for radio pulsations by Gorham et al. (1996)
at 430 and 1420 MHz.  No pulsations were observed above a level of 0.7 mJy
at 1420 MHz.  (This is rather weak support of the ``radio-quiet''
classification since a large fraction of pulsars
discovered in modern surveys, (e.g. Manchester et al 2001) have lower 
fluxes than this limit.)

The X-ray spectra of all these objects are close to blackbody spectra.  The
classical blackbody model, however, predicts a temperature which is
too high and a surface area which is too small when compared
with generally accepted models of neutron stars.  For example, 
Chakrabarty et
al. (2001) have fit the spectrum of the Cas A object and find a
temperature of 0.49 keV and star radius of 0.52 km; compared with
the 8-16 km radius expected.  The radiation source could be a single
hot spot on the star surface but this is hard to reconcile with the
observed lack of pulsations ($\leq25\%$ pulsed) determined by Murray et
al. (2001). 

A light-element atmosphere will reduce the derived temperature
 and increase the derived radius.  Lloyd, Hernquist \& Heyl (2002)
have calculated emergent spectra and find that the actual temperature,
$T_{eff}$, is always less than the temperature, $T_{bb}$, derived by
fitting a blackbody spectrum to the observed data.  Chakrabarty et
 al. (2000) find that a H atmosphere model applied to
 the Cas A source yields a temperature of 0.26 kev and a radius of 2.2
 km.  They conclude that existing surface-radiation
 models do not explain the Cas A object and that accretion models also
 fail to account for the lack of an optical counterpart.  The more 
luminous Kes 79 object, if a classical blackbody, would require a 
 radius of only 1.0 km to achieve the observed luminosity at
the measured temperature of $T_{bb}$ = 0.48 kev.  Application of the
light atmosphere model of Lloyd, Hernquist and Heyl predicts that
$T_{eff}$ is a factor of 1.8 lower than $T_{bb}$, or 0.27 keV.  
This lower temperature would increase the required
 radius to 3.2 km, still short of the 8-16 km expected.  Since the
theoretical models used so far have been simple, perhaps adjustments
might be made in the model atmosphere to achieve a fit with a
standard radius.  Obviously better models are needed. 

Among the 5 Central Compact Objects (CCOs) mentioned, The most luminous is
in the largest remnant, G347.3-0.5.  The source in Kes 79 is the
second most luminous.  Kes 79 is about the same size as PKS 1209-51/52
but in a denser environment, so is probably older.  Thus the central
source in Kes 79 may also be the second oldest specimen of the 
``radio-quiet'' isolated pulsar group.  This agrees exactly with 
the conclusion of
Pavlov, Sandwal, \& Garmire (2002) that the older CCOs are more luminous.  
If these central objects are similar, they are certainly not cooling 
rapidly and the apparent increase in emitting area with age is a mystery.

Certainly, the source in Kes 79 is worthy of more
study.  It happens to be in a remnant which has a well defined outer
shell and, interpreting this as a shock, one can derive
 information about the SN explosion
which produced it.  Future work should include sensitive 
searches for radio and/or X-ray pulsations and, although none is
expected, a  sensitive search for an optical counterpart.

This work was supported by NASA Grant GO1-2067X. \\



\begin{thebibliography}{}

\bibitem[Chakrabarty et al. 2001]{CH01} Chakrabarty, D., Pivovaroff,
M,J., Hernquist, L.E., Heyl, J.S. \& Narayan, R. 2001, ApJ 548, 800

\bibitem[(CXC 2001)]{CX01}Chandra X-ray Center, 2001, Proposers
Observatory Guide Rev. 4.0, TD 403.00.004

\bibitem[Frail \& Clifton (1989)]{FRCL89} Frail, D.A. \& Clifton T.R.
1989, ApJ 336, 854.

\bibitem[(Green et al 2002)]{WI02} Green P., et al. 2002, in preparation

\bibitem[(Grindlay et al. 2002)]{GR02} Grindlay, J., et al. 2002, in preparation

\bibitem[Gorham et al. (1996)]{GO96} Gorham, P.W., Ray, P.S.,
Anderson, S.B., Kulkarni, S.R. \& Prince, T.A. 1996, ApJ 458, 257

\bibitem[]{329} Han, J.L. 1997, A\&A 318, 485
 
\bibitem[]{}Helfand, D.J., Gotthelf, E.V., \& Halpern, J.P. 2001, ApJ
556, 380
 
\bibitem[]{331} Space Telescope Science Institute, Guide Star Catalog version 2.2, 2001, http://www-gsss.stsci.edu/gsc/gsc2/GSC2home.htm

\bibitem[Kesteven (1968)]{KE68} Kesteven, M.J.L. 1968, Aust. J. Phys. 21, 369.

\bibitem[]{LHH02} Lloyd, D.A., Hernquist, L. \& Heyl, J.S. 2002, in preparation
\bibitem[(Manchester et al 200)1]{MA01} Manchester, R.N., et al 2001, MNRAS
328, 17

\bibitem[(Murray et al, 2001)]{MU01} Murray, S.S., Ransom, S.M., Juda,
M., Hwang, U. \& Holt, S.S. 2001, Astro-ph 0106516

\bibitem[Pavlov et al. 2002]{PA020} Pavlov, G.G., Sanwal, D., Garmire, 
G.P. \& Zavlin, V.E. 2002, Astro-ph/0112322, to appear in ``Neutron 
Stars and Supernova Remnants'' eds. P.O. Slane \& B.M. Gaensler

\bibitem[Pavlov et al. 2002]{PA02} Pavlov, G.G., Zavlin, V.E., Sanwal,
D. \& Trumper, J. 2002, ApJ 569, L95

\bibitem[(Seaquist \& Gilmore 1982)]{SG82} Seaquist, E.R. \& Gilmore,
W.S. 1982, AJ 87, 378. 

\bibitem[(Seward \& Velusamy, 1995)]{SV95} Seward, F.D. \& Velusamy,
T. 1995, ApJ 439, 715. 

\bibitem[Slane et al 1999]{SL99} Slane, P.O., Gaensler, B.M., Dame,
T., Hughes, J.P., Plucinsky, P.P. \& Green, A. 1999, ApJ 525, 357

\bibitem[Slane et al 2000]{SL00} Slane, P.O., Chen, Y., Schulz, N.,
Seward, F., Hughes, J.P., \& Gaensler, B. 2000, ApJ 533, L29

\bibitem[(Sun et al. 2000)]{SU02} Sun, M., Seward, F.D., Slane, P.O. \&
Smith, R.K. 2002, In preparation.

\bibitem[(Sun \& Wang, 2000)]{SW00} Sun, M. \& Wang, Z-R. 2000,
Adv. Space Res. 25, No. 3/4, 549

\bibitem[(Toor \& Seward, 1977)]{TO77} Toor, A., \& Seward, F.D. 1977,
ApJ 216, 560.

\bibitem[(Velusami, Becker \& Seward 1991)]{VBS91} Velusami, T.,
Becker, R.H. \& Seward, F.D. 1991, AJ 102, 676.

\bibitem[(Zavlin et al. 2000)]{ZA00} Zavlin, V., Pavlov, G., Sanewal,
D. \& Truemper, J. 2000, ApJ 540, L25.

\bibitem[(Zavlin et al. 1999)]{ZA99} Zavlin, V., Truemper, J. \& Pavlov,
G. 1999, ApJ 525, 959.

\end{thebibliography}
\end{document}